\documentstyle[aps,12pt,epsfig,manuscript,amsbsy]{revtex}
\global\firstfigfalse
\begin{document}
\preprint{INJE-TP-99-7}
\def\overlay#1#2{\setbox0=\hbox{#1}\setbox1=\hbox to \wd0{\hss #2\hss}#1%
\hskip -2\wd0\copy1}

\title{Noncommutative spacetime and the fractional quantum Hall effect}

\author{ Y.S. Myung and H.W. Lee}
\address{Department of Physics, Inje University, Kimhae 621-749, Korea}

\maketitle

\begin{abstract}
We propose the two formalisms for obtaining the noncommutative spacetime 
in a magnetic field. 
One is the first-order formalism and the other is the second-order formalism.
Although the noncommutative spacetime is realized manifestly in the 
first-order formalism, the second-order formalism would be more useful 
for calculating the physical quantities in the noncommutative geometry than 
the first-order one.
Several interesting points for string theory and fractional 
quantum Hall effect are discussed.
In particular, we point out that the noncommutative geometry is closely 
related to the fractional quantum Hall effect(FQHE).
\end{abstract}

\newpage
Recently noncommutative geometry has attracted nuch interest in string 
and M-theory in the $B$-field\cite{Con98JHEP02003,Dou98JHEP02008,She99PLB119,Mal9908134,Sei9908142,Big9908056}.
Gauge theory on noncommutative boundary space are relevant to the 
quantization of D-branes in $B_{\mu\nu}$ fields.
It is very important to notice that the noncommutativity can be easily 
realized by turning on a magnetic field in the plane.
This situation is also applicable to studying the FQHE\cite{Gir9907002}.

In this letter we study the two models which give us the same 
noncommutative effect.
It is well known that the quantum mechanics of a nonrelativistic 
particle in a constant magnetic field(${\rm\bf B}=B\hat{\rm\bf k}$) produces 
the noncommutative momentum $\boldsymbol{\pi}$.
Hence we start with the second-order Lagrangian\cite{Dun93NPB114,Myu9910083}
\begin{equation}
{\cal L}_m = {1 \over 2} m \dot{\rm\bf r}^2 +
  { q \over c} \dot{\rm\bf r} \cdot {\rm\bf A} - V( r)
\label{second-lagrangian}
\end{equation}
where ${\rm\bf r} = (x,y)$ is the coordinates and the vector 
potential
($\boldsymbol{\nabla} \times {\rm\bf A} = {\rm\bf B}$)
is given by $A^i = - \epsilon^{ij} r^j B/2$.
The corresponding Hamiltonian is given by
\begin{equation}
{\cal H}_m = {1 \over { 2m}} \left ( {\rm\bf p} - {q \over c} {\rm\bf A} 
   \right )^2 + V( r).
\label{second-hamiltonian}
\end{equation}
Here we have two momenta (${\rm\bf p}, \boldsymbol{\pi} =
m \dot{\rm\bf r} = {\rm\bf p} - {q \over c} {\rm\bf A}$).
${\rm\bf p}$ is the canonical momentum which is a gauge variant quantity 
and thus is not a physical observable.
On the other hand $\boldsymbol{\pi}$ is the mechanical momentum which is 
a gauge invariant quantity and thus is a physical observable.
Now let us calculate the commutator of $\boldsymbol{\pi}$
\begin{equation}
\left [ \pi^i, \pi^j \right ] = i B \epsilon^{ij}
\label{pi-commutator}
\end{equation}
with $q/c=-1$.
This shows that $\boldsymbol{\pi}$ becomes a noncommutative momentum 
when a magnetic field is turned on.
However, one finds the commutative spacetime such as 
$[ r^i, r^j ] = 0$.

In order to obtain the noncommutative geometry, 
we take the $m \to 0$ limit\cite{Dun93NPB114}
\begin{equation}
{\cal L}_{m\to 0} = - { B \over 2} \dot r^i \epsilon_{ij} r^j -V(r).
\label{first-lagrangian}
\end{equation}
This is the first-order Lagrangian and thus, has a sympletic structure
which enforces the noncommutative relation\cite{Fad88PRL1692}
\begin{equation}
 [ r^i, r^j ] = {i \over B} \epsilon^{ij}.
\label{r-commutator}
\end{equation}
The corresponding Hamiltonian ${\cal H}_{m\to 0}$ is 
solely given by the potential 
\begin{equation}
{\cal H}_{m\to 0} = {{\partial {\cal L}_{m\to 0} } \over 
  {\partial \dot{\rm\bf r}}} \cdot \dot{\rm\bf r} -{\cal L}_{m\to 0} 
  = V( r).
\label{first-hamiltonian}
\end{equation}
This can be also derived from (\ref{second-hamiltonian}) by imposing 
the constraint
\begin{equation}
\boldsymbol{\pi} = m \dot{\rm\bf r} = {\rm\bf p} + {\rm\bf A} \simeq 0.
\label{constraint}
\end{equation}
Actually (\ref{r-commutator}) emerges from 
obtaining the same equation of motion 
$\dot r^i = - {\epsilon^{ij} \over B} {{\partial V} \over {\partial r^j}}$
by using both the Lagrangian formalism of (\ref{first-lagrangian}) 
and the Hamiltonian framework of (\ref{first-hamiltonian}).

We wish to comment the following points:
\begin{enumerate}
\item We note that in the limit of $m\to 0$, the phase space of four 
($p_x, p_y, x, y$) is reduced to two ($x,y$). 
This is so because of a consequence of the constraint (\ref{constraint}).
This implies that there is a reduction of degrees of freedom in the 
noncommutative spacetime, compared to the ordinary case.
Hence we expect that this leads to a negative 
entropy correction in noncommutative super 
Yang-Mills theory\cite{Cai9910092}.
\item The limit of $m\to 0$ with finite $B$ is actually a projection onto 
the Lowest Landau Level(LLL).
In other words, the LLL can be singled out by 
setting $m\to 0$\cite{Dun93NPB114}.
The same projection is also performed by the limit of $B\to \infty$ with 
finite $m$\cite{Myu9910083,Vei92NPB715}.
\item It is difficult to compute some physical quantities with 
the constraint system (${\cal L}_{m\to 0}, {\cal H}_{m\to 0}$).
However, a similar computation with (${\cal L}_m, {\cal H}_m$) may 
be straightfoward because of the simpler, expanded sympletic structure.
It was argued that if an operator ${\cal O}$ commutes 
with the constraint (\ref{constraint}) 
when the latter vanishes, the results in two approaches 
will be the same\cite{Fad88PRL1692}.
Hence it would be correct that starting with the second-order formalism
${\cal H}_m$, and then one requires either the limit of 
$m\to 0$ or the limit of $B \to \infty$ to obtain
the effect of noncommutative geometry.
Our previous calculation was performed along 
this line\cite{Myu9910083,Kim93PRD4839}.
The limiting procedure was carried out through the action of
$\omega_c = B/2m \to \infty$ at the final stage of calculation.
For a dipole configuration with the harmonic interaction\cite{Big9908056}, 
it would be better to use the second-order 
formalism than the first-order formalism.
\item For the string(gauge) theory calculations\cite{Big9908056,She9901080}, 
we propose a new procedure 
with the second-order formalism to 
obtain the noncommutative effect, since the calculation on 
noncommutative spacetime is not easy.
\item The noncommutative spacetime (\ref{r-commutator}) 
leads to the FQHE.

Using the second-quantized formalism, one finds the Hamiltonian
${\cal H}_0$
\begin{equation}
{\cal H}_0 = { 1\over 2m} \psi^\dag 
\left ({\rm\bf p} + {\rm\bf A} \right )^2 \psi.
\label{h0}
\end{equation} 
This Hamiltonian is a conerstone to study the FQHE\cite{Iso92PLB123}.
The limit of $m\to 0$ leads to $\left ({\rm\bf p} + {\rm\bf A} \right )
\psi_{\rm LLL} =0$, which means that the Fermi particles reside in the LLL.
A general solution to this operator constraint is given by
\begin{equation}
\psi_{\rm LLL}(z, \bar z ) = \sum_{n=0}^\infty c_n \varphi_n(z, \bar z ),
\label{lll-solution}
\end{equation}
where $z = \sqrt{B/2 } ( x + i y)$, $\bar z = \sqrt{B/2 } ( x - i y)$, 
$\{ c_n, c_m^\dag \} = \delta_{nm}$.
$\varphi_n$ is the $n$-th single particle state in the LLL,
\begin{equation}
\varphi_n(z, \bar z ) = { 1 \over \sqrt{\pi n !}} z^n 
 \exp\left ( - {1 \over 2} | z |^2 \right )
\label{varphin}
\end{equation}
with $B = 2$ for convenience.
Since the LLL wave functions $\{\varphi_n \}$ are incomplete in the view of 
the total Hilbert space (full Landau levels), the fields have the 
unconventional commutator\cite{Mar93IJMPB4389}
\begin{equation}
\{ \psi^\dag_{\rm LLL}(z_1, \bar z_1 ), \psi_{\rm LLL}(z_2, \bar z_2 ) \}
={1 \over \pi} \exp \left ( - {1 \over 2} | z_1 - z_2 |^2 
   + {1 \over 2} ( \bar z_1 z_2 - \bar z_2 z_1 ) \right )
\equiv \left \{ z_1 | z_2 \right \}
\label{equal-time}
\end{equation}
instead of a conventional form of 
$\{ \psi^\dag(z_1, \bar z_1 ), \psi(z_2, \bar z_2 ) \} = 
\delta^2(z_1 - z_2 )$.
That is, a bilocal kernel $\{ z_1 | z_2 \}$ is introduced as 
a LLL analogue of the 
delta function and retains its reproducing property
\begin{equation}
\int d^2 z_1 F(z_1) \{ z_1 | z_2 \} = F(z_2)
\label{reproduce}
\end{equation}
where $F(z)$ is any function of the form $F(z) = f(z) e^{-|z|^2/2}$.
The field $\psi_{\rm LLL}$ takes this form and thus one finds
\begin{equation}
\psi_{\rm LLL} (z, \bar z) = \int d^2 z \psi(z', \bar z') \{ z' | z \},
\label{psiLLL}
\end{equation}
which means the overcompleteness of the LLL.
In other words, we cannot vary $\psi_{\rm LLL}(z, \bar z)$ 
independently at different point $(z, z')$ in the LLL, since 
neighboring fields are linked by (\ref{psiLLL}).
Actually this overcompleteness comes from $[z, \bar z ] =1 $(
another form of (\ref{r-commutator}))
and thus it is related to the noncommutative space.

We wish to describe the many-electron state out of any antisymmetric 
holomorphic function
\begin{equation}
| f \rangle = \int \left [ \prod_{k=1}^N d^2 z_k \right ] 
    f(z_1, z_2, \cdots, z_N ) 
 e^{-{1 \over 2} \sum_{k=1}^N | z_k |^2 }
  \psi^\dag_{\rm LLL}(z_1) \psi^\dag_{\rm LLL} (z_2) 
\cdots \psi^\dag_{\rm LLL}(z_N) | 0 \rangle. 
\label{holomorphic}
\end{equation}
Thanks to the anticommutation relation (\ref{equal-time}) and the 
reproducing kernel (\ref{reproduce}), one enables to to express 
the inner product of two such states as
\begin{equation}
\langle f | g \rangle = N! 
\int \left [ \prod_{k=1}^N d^2 z_k \right ]
\overline{f(z_k)} g(z_k) 
e^{-\sum_{k=1}^N | z_k |^2 }.
\label{innerproduct}
\end{equation}
In particular, defining the state
\begin{equation}
 | z_1, z_2, \cdots z_N \rangle = 
\psi^\dag_{\rm LLL} (z_1) \psi^\dag_{\rm LLL} (z_2) \cdots 
\psi^\dag_{\rm LLL}(z_N) | 0 \rangle ,
\label{manystate}
\end{equation}
we recover the many-electron wavefunction
\begin{equation}
\langle z_1, z_2, \cdots z_N | f \rangle =
  f(z_1, z_2, \cdots, z_N ) 
  e^{-{1 \over 2} \sum_{k=1}^N | z_k | ^2 }.
\label{manyelectron}
\end{equation}
Finally we obtain the Laughlin wavefunction which is essential to the 
FQHE by taking $f = \prod_{k<l} (z_k - z_l )^{2n+1}$.
The key physics of the FQHE results from the rearrangement of the 
nearly degenerate states$\{\varphi_n \}$ in the LLL.
Here the noncommutative feature of $[z, \bar z]=1$
plays an important role in constructing the holomorphic 
eigenfunction $\{\varphi_n \}$ of ${\cal H}_{m\to 0}$ in 
(\ref{first-hamiltonian})\cite{Dun93NPB114}.
Hence we insist that although the FQHE is an effect of many-body 
interaction,
the noncommutative spacetime leads to the FQHE.
\item In order to get the Moyal bracket, we introduce the charge density 
operator $\hat \rho(z) = \psi^\dag_{\rm LLL}(z) \psi_{\rm LLL}(z)$.
And the second-quantized Hamiltonian ${\cal H} = {\cal V}$ is given by
\begin{equation}
{\cal V} = \int d^2 z \psi^\dag_{\rm LLL} V(z, \bar z) \psi_{\rm LLL}(z) .
\label{potentialV}
\end{equation}
We commute ${\cal H}$ through $\psi_{\rm LLL}$ to find
\begin{equation}
i \partial_t \psi_{\rm LLL} (z) = 
\int d^2 z' V(z') \psi_{\rm LLL}(z') \{ z' | z \} .
\label{partialt}
\end{equation}
Introducing an apodized potential
\begin{equation}
\widetilde V (z_1) = \int {d^2 z \over \pi }
e^{- | z - z_1 | ^2 } V(z) ,
\label{apodized}
\end{equation}
we derive an analogue of the Moyal bracket
\begin{equation}
\partial_t \hat \rho (z) = {1 \over i } \sum_{n=1}^\infty 
{1 \over n! } \left \{ \partial_{\bar z}^n \widetilde V (z, \bar z) 
\partial_z^n \hat \rho (z) - 
\partial_{z}^n \widetilde V (z, \bar z) 
\partial_{\bar z}^n \hat \rho (z)
\right \}.
\label{moyal}
\end{equation}
\end{enumerate}

In conclusion, we propose two formalisms which gives us the same 
noncommutative effect.
In particular, we show that the noncommutative geometry is closely 
related to the FQHE.

\section*{Acknowledgement}
We are greateful to R. Jackiw for sending Ref.\cite{Dun93NPB114} 
to us and G.W. Kang for helpful discussions.
This work was supported by the Brain Korea 21
Program, Ministry of Education, Project No. D-0025.

\end{document}